\documentclass[conference]{IEEEtran}
\IEEEoverridecommandlockouts
\usepackage{cite}
\usepackage{amsmath,amssymb,amsfonts}
\usepackage{algorithmic}
\usepackage{graphicx}
\usepackage{adjustbox}
\usepackage{booktabs}
\usepackage{textcomp}
\usepackage{xcolor}
\def\BibTeX{{\rm B\kern-.05em{\sc i\kern-.025em b}\kern-.08em
    T\kern-.1667em\lower.7ex\hbox{E}\kern-.125emX}}
\begin{document}
\title{Multi-Loss Convolutional Network with Time-Frequency Attention for Speech Enhancement\\
}

\author{\IEEEauthorblockN{1\textsuperscript{st} Liang Wan}
\IEEEauthorblockA{\textit{School of Communication and Information Engineering}\\
\textit{Chongqing University of Posts and Telecommunications}\\
Chongqing, China \\
Email:s210131221@stu.cqupt.edu.cn}
\and
\IEEEauthorblockN{2\textsuperscript{nd} Hongqing Liu}
\IEEEauthorblockA{\textit{School of Communication and Information Engineering}\\
\textit{Chongqing University of Posts and Telecommunications}\\
Chongqing, China \\
Email:hongqingliu@outlook.com}
\and
\IEEEauthorblockN{3\textsuperscript{rd} Yi Zhou}
\IEEEauthorblockA{\textit{School of Communication and Information Engineering}\\
\textit{Chongqing University of Posts and Telecommunications}\\
Chongqing, China \\
Email:zhouy@cqupt.edu.cn
}
\and
\IEEEauthorblockN{4\textsuperscript{rd} Jie Jia}
\IEEEauthorblockA{
\textit{vivo Mobile Commun co Ltd}\\
China \\
Email:jie.jia@vivo.com
}
}

\maketitle

\begin{abstract}
The Dual-Path Convolution Recurrent Network (DPCRN) was proposed to effectively exploit time-frequency domain information. By combining the DPRNN module with Convolution Recurrent Network (CRN), the DPCRN obtained a promising performance in speech separation with a limited model size. In this paper, we explore self-attention in the DPCRN module and design a model called Multi-Loss Convolutional Network with Time-Frequency Attention(MNTFA) for speech enhancement. We use self-attention modules to exploit the long-time information, where the intra-chunk self-attentions are used to model the spectrum pattern and the inter-chunk self-attention are used to model the dependence between consecutive frames. Compared to DPRNN, axial self-attention greatly reduces the need for memory and computation, which is more suitable for long sequences of speech signals. In addition, we propose a joint training method of a multi-resolution STFT loss and a WavLM loss using a pre-trained WavLM network. Experiments show that with only 0.23M parameters, the proposed model achieves a better performance than DPCRN.
\end{abstract}


\section{Introduction}
Excessive noise and reverberation can significantly impair the performance of automatic speech recognition (ASR) systems and reduce speech intelligibility during communication. To improve speech intelligibility and perceptual quality, speech enhancement techniques are generally employed to separate clean speech from background interferences. This separation helps eliminate the effects of noise and reverberation, resulting in clearer and more understandable speech.

Deep neural networks (DNNs) have become increasingly popular in audio processing applications, and have been shown to be highly effective in speech enhancement tasks. In terms of domain used, DNN-based speech enhancement methods can be broadly divided into two categories: time-frequency (TF) domain\cite{20061486,6639038,7472934,6665000,Yin_Luo_Xiong_Zeng_2020} and time domain\cite{fu2018end,Wave-U-Net,zhang2020furcanext} methods. Methods operating in the time domain directly predict clean speech waveforms by end-to-end training, avoiding the challenge of estimating phase information in the TF domain. Conv-Tasnet\cite{luo2019conv-tasnet}, as a representative time-domain method, employs a 1-D convolutional neural network to encode the waveform into effective representations for accurate speech estimation, and then recovers the waveform by a decoder consisting of transposed convolutional layers. However, modeling extremely long sequences is challenging in the time domain, hence deep convolutional layers such as wave-u-net\cite{Wave-U-Net} are often needed for feature compression.
TF-domain methods operate on the spectrogram, under the assumption that the detailed structures of speech and noise can be more easily distinguished and separated through time-frequency representations. In TF-domain methods, training targets can be broadly categorized into two groups: masking-based targets and mapping-based targets. Masking-based targets such as ideal binary mask (IBM)\cite{wang2005ideal}, ideal ratio mask (IRM)\cite{6639038}, and spectral magnitude mask (SMM)\cite{wang2014training}, only consider the magnitude differences between clean speech and mixture speech while ignoring phase information. Subsequently, complex ratio mask (CRM)\cite{8682834} was proposed as a masking-based target, which enhances both the real and imaginary components of the division between clean speech and mixture speech spectrograms, allowing for a better speech reconstruction.

For TF domain method, Dual-Path Convolution Recurrent Network (DPCRN)\cite{le2021dpcrn} combining DPRNN\cite{luo2020dual} and CRN\cite{crn2018}, it is possible to obtain a well-behaved model. On the basis of DPCRN, in this work, a new model called Multi-Loss Convolutional Network with Time-Frequency Attention(MNTFA) for speech enhancement is proposed. It replace DPRNN with an Axial Self-Attention (ASA)\cite{zhang2022multi} block consisting of two ASAs. ASA improves network ability to capture long-range relations between features in time domain and frequency domain, respectively. Compared to DPRNN, ASA greatly reduces the need for memory and computation, which is more suitable for long sequences of speech signals. The high-level features extracted by encoder are fed into ASA blocks for further processing. Subsequently, the decoder reconstructs the low-resolution features to the output.

While DNN-based speech enhancement has made rapid progress, it still faces challenges in real-world applications due to low signal-to-noise ratio (SNR), high reverberation, and far-field pickup. These issues remain significant obstacles to achieving high-quality speech enhancement in practical settings. Previous studies have found that blindly feeding the output of a speech enhancement system (SE) to an ASR system can result in degraded performance. However, incorporating a pre-trained ASR system during the SE model training stage has been shown to improve ASR performance. For the proposed model, we employ a multi-loss strategy formed by SE loss and an ASR loss to simultaneously ensure good SE and ASR performance. The SE loss consists by a mean square error (MSE) loss for noise suppression and multi-resolution STFT loss\cite{yamamoto2020parallel} for speech preservation. The ASR loss\cite{hsieh2020improving} is formed by a pretrained WavLM\cite{chen2022wavlm} network.

The proposed system is evaluated using PESQ and STOI for SE performance and word error rate (WER) for ASR performance. On the test datasets, our model achieves competitive results as DPCRN while having a smaller number of parameters and computational complexity.

\begin{figure*}[h]
	\centering
	\includegraphics [scale=0.5]{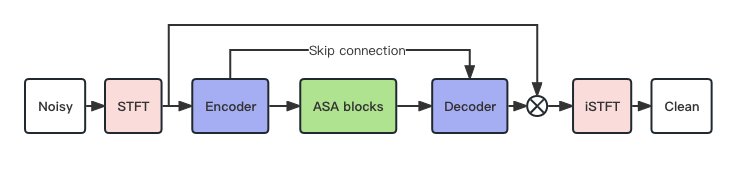}
	\caption{ Architecture of the proposed MNTFA}
	\label{figure1}
\end{figure*}

\section{Network}
\subsection{Problem formulation}

The observed noisy speech can be represented as a sum of the clean speech and noise signals in the time domain, which is $y(t) = s(t) + n(t)$, where $y(t)$, $s(t)$, and $n(t)$ refer to the noisy, clean, and noise signals, respectively. The equation can be converted to the time-frequency domain by utilizing the Short-Time Fourier Transform (STFT), given by
\begin{equation}
Y(t, f) = S(t, f)+N(t, f),
\label{eq1}
\end{equation}
where $Y(t, f)$, $S(t, f)$, and $N(t, f)$ represent the time-frequency bins of the noisy, clean, and noise speech spectrograms, respectively, at time frame $t$ and frequency index $f$. To recover clean speech from a noisy mixture, a mask $M(t, f)$ is typically estimated and multiplied with the noisy speech spectrogram $Y(t, f)$\cite{wang2018supervised}.
To better utilize phase information, we adopt complex ratio mask (CRM), denoted by $M(t, f) =M_r(t, f) + iM_i(t, f)$, where $M_r(t, f)$ and $M_i(t, f)$ represent the real and imaginary parts of the mask. CRM better reconstructs speech by enhancing both real and imaginary spectrogram\cite{hu2020dccrn} simultaneously. The denoising process can be represented as the complex product of mask and noisy speech in the form of:
\begin{equation}
\widetilde{S}(t, f) = Y(t, f)\bigodot M(t, f),
\label{eq2}
\end{equation}
where $\bigodot$ denotes element-wise multiplication and $\widetilde{S}(t, f)$ is the enhanced speech.

\subsection{Model architecture}
The Dual-Path Convolution Recurrent Network combining DPRNN and CRN achieves a state-of-the-art (SOTA) performance in single-channel speech separation task. Similar to the original DPRNN, DPCRN consists of an encoder, a dual-path RNN module, and a decoder. The structure of the encoder and decoder is similar to CRN. DPCRN replaces the RNN part of the CRN with the DPRNN modules. It is believed that modeling the dependence of frequency is beneficial to speech enhancement due to the harmonic structures of speech. Instead of learning the dependence in the time domain, the intra-chunk RNNs are applied to model the spectral patterns in a single frame and inter-chunk RNNs model the time dependence of a certain frequency.

\begin{figure}[htbp]
	\centering
	\includegraphics [scale=0.4]{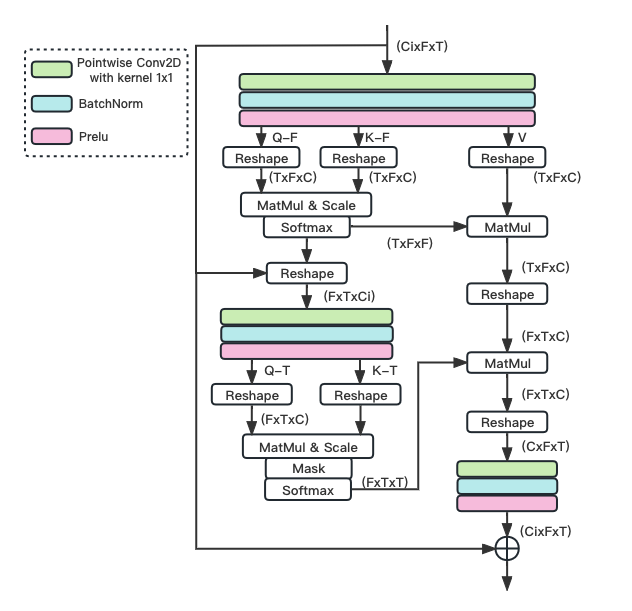}
	\caption{Axial self attention module}
	\label{figure2}
\end{figure}

\begin{figure}[htbp]
	\centering
	\includegraphics [scale=0.35]{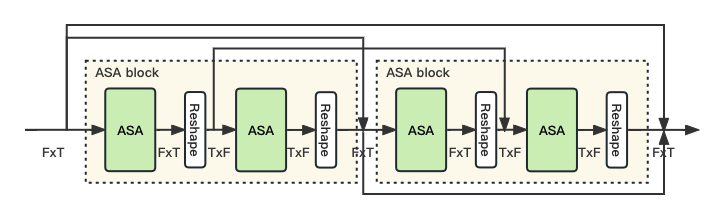}
	\caption{ASA blocks}
	\label{figure3}
\end{figure}

However, the DPCRN suffers from a high computational complexity. In this paper, we propose a 
 novel model with an Axial Self-Attention block consisting of two ASAs to model long-time dependencies. Figure \ref{figure1} shows the overall network structure of the proposed model. Axial Self-Attention (ASA) was introduced in \cite{zhang2022multi} as a more efficient approach that reduces the demand for memory and computation resources. ASA is particularly useful for processing lengthy speech signal sequences. The ASA structure, illustrated in Figure \ref{figure2}, consists of input channels represented by $C_i$ and attention channels represented by $C$. ASA calculates attention scores along the frequency and time axis, which are referred to as F-attention and T-attention, respectively. These attention score matrices enhance the model's ability to capture long-range dependencies in both the time and frequency domains simultaneously. Similar to DPCRN, we construct an ASA block by combining two ASAs. Figure \ref{figure3} shows the structure of ASA block, and as shown in figure \ref{figure3}, the input and output dimensions of the first ASA are T, F. We rearrange the output dimensions of the first ASA to F, T, and pass it to the second ASA. In MNTFA, we use two ASA blocks to capture long-range relations between features. To alleviate the problem of gradient vanishing, a residual connection\cite{he2016deep} was utilized between the intra-chunk ASA and inter-chunk ASA. We found that this development enables the model to better learn both temporal and spectral contextual information simultaneously, which is particularly relevant in the context of speech processing.

\subsection{Training target}
When training, MNTFA estimates CRM and is optimized by signal approximation (SA)\cite{he2015delving}. Given the complex-valued STFT spectrogram of clean speech S and noisy speech $Y$, CRM is defined as
\begin{equation}
CRM = \frac{Y_rS_r+Y_iS_i}{{Y_r}^2+{Y_i}^2}+j\frac{Y_rS_i-Y_iS_r}{{Y_r}^2+{Y_i}^2},
\label{eq3}
\end{equation}
where $Y_r$ and $Y_i$ denote the real and imaginary parts of the noisy complex spectrogram, respectively. The real and imaginary parts of the clean complex spectrogram are represented by $S_r$ and $S_i$. By multiplying the spectrogram of noisy speech $Y = Y_r + Y_i$ with the estimated mask $M = M_r + M_i$, we obtain the enhanced spectrogram in the form of: $\widetilde{S} = Y_rM_r – Y_iM_i + i(Y_rM_i + X_iM_r)$, which is converted back to the waveform using inverse STFT (iSTFT), given by
\begin{equation}
\widetilde{s} = iSTFT(\widetilde{S} ).
\label{eq4}
\end{equation}

\subsection{Loss function}
We use three loss functions to measure the estimation error from the signal, perceptual, and ASR aspects, respectively.
To suppress the noise for the low-SNR time-frequency bins, we employ a metric called mean square error (MSE) of the spectrogram, which is
\begin{equation}
\begin{split}
L_{MSE}= & \log(MSE(S_r,\widetilde{S_r} )+MSE(S_i,\widetilde{S_i} ) \\
& +MSE(|S|,|\widetilde{S}| )).
\end{split}
\label{eq5}
\end{equation}
The MSE loss is composed of three components that individually quantify the differences in the real part, imaginary part, and magnitude between the predicted spectrogram and the ground truth.

To enhance the clarity and intelligibility of speech following the denoising process, we used a multi-resolution STFT auxiliary loss\cite{yamamoto2020parallel}. Similar to prior research\cite{yamamoto2019probability}, we define a single Short-Time Fourier Transform (STFT) loss as
\begin{equation}
L_s =  L_{sc}(s,\hat{s})+L_{mag}(s,\hat{s}),
\label{eq6}
\end{equation}
where $s$, $\hat{s}$ are the clean and estimated time-domain waveform. $L_{sc}$ and $L_{mag}$ denote spectral convergence and log STFT magnitude loss, respectively, which are \cite{arik2018fast}

\begin{equation}
L_{sc}(s,\hat{s}) = \frac{\lVert \left|STFT(s)\right|-\left|STFT(\hat{s})\right| \rVert_F}{\lVert \left|STFT(s)\right|\rVert_F},
\label{eq7}
\end{equation}

\begin{equation}
L_{mag}(s,\hat{s}) = \frac{1}{N}\lVert \log \left|STFT(s)\right|-\log \left|STFT(\hat{s})\right| \rVert_1,
\label{eq8}
\end{equation}
where $\lVert .\rVert_F$ and $\lVert .\rVert_1$ denote the Frobenius and $L_1$ norms, respectively; $|STFT(.)|$ and N denote the STFT magnitudes and number of elements in the magnitude, respectively.
The multi-resolution STFT loss is obtained by summing the STFT losses computed with different analysis parameters such as FFT size, window size, and frame shift. If there are $M$ STFT losses, the multi-resolution STFT auxiliary loss ($L_{aux}$) is
\begin{equation}
L_{aux} = \frac{1}{M}\sum_{m=1}^ML_s^{(m)}.
\label{eq9}
\end{equation}
When using STFT for time-frequency representation of signals, the balance between time and frequency resolution is crucial. Increasing the window size improves the frequency resolution but reducing the temporal resolution. By using multiple STFT losses with different analysis parameters, the model can better learn the time-frequency characteristics of speech and avoid becoming overfit to a fixed STFT representation, which could result in suboptimal performance in the waveform-domain.

Moreover, an ASR-oriented loss $L_{ASR} $ is used to reduce the speech distortion of enhanced speech and reduce WER for a given ASR system. More specifically, KL divergence distance between clean speech ASR embedding and estimated speech ASR embedding is employed as our ASR loss. The ASR embeddings are extracted by the pre-trained WavLM.

Our final loss function for the model is defined as a linear combination of the MSE loss, Multi-resolution STFT auxiliary loss, and ASR loss, with a equal weighting, given by
\begin{equation}
L = L_{MSE}+L_{aux} + L_{ASR},
\label{eq10}
\end{equation}

\begin{table}[htbp]
\caption{PESQs and STOI of different methods.}
\centering
\begin{adjustbox}{scale=0.95}
\begin{tabular}{@{}ccccccccc@{}}
\toprule
Metrics & \multicolumn{4}{c}{PESQ}  & \multicolumn{4}{c}{STOI (\%)} \\ \midrule
SNR(dB) & -5   & 0    & 5    & Avg. & -5    & 0     & 5     & Avg.  \\ \midrule
Noisy   & 1.58 & 1.84 & 2.14 & 1.85 & 65.50 & 73.25 & 79.53 & 72.76 \\
DCCRN   & 2.30 & 2.68 & 3.02 & 2.67 & 74.52 & 81.32 & 86.53 & 80.79 \\
DPCRN   & 2.36 & 2.67 & 2.91 & 2.65 & 75.08 & 81.60 & 86.28 & 80.98 \\
MNTFA   & 2.42 & 2.72 & 3.02 & 2.72 & 75.64 & 81.97 & 86.78 & 81.46 \\ \bottomrule
\end{tabular}
\end{adjustbox}
\label{tbl:table1}
\end{table}

\section{Experiments}

\subsection{Datasets}\label{AA}
The clean speech and noise datasets from the DNS Challenge (ICASSP 2022)\cite{reddy2020interspeech}  are used for our experiments. By randomly pairing speech and noise signals, we synthesize a training set with 500 hours samples and a validation set with 50 hours, in both of which the signal-to-noise ratio (SNR) is randomly sampled between -5 dB and 5 dB. Note that all speech and noise signals are randomly truncated to 10 seconds before mixing. We additionally use the synthetic test set released by DNS Challenge for testing. All the audios used is sampled at 16 kHz.
\begin{table}[htbp]
\caption{The model size, MACS and WER of different models.}
\centering
\begin{tabular}{cccc}
\hline
Model & Para.(M) & MACs(GFLOPS)  & WER \\ \hline
Noisy &           &              & 13.12     \\
DCCRN & 3.67     & 14.38       & 15.23  \\
DPCRN & 0.8      & 8.45       & 13.71 \\
MNTFA & 0.23     & 1.89      & 12.55  \\ \hline
\end{tabular}
\label{tbl:table2}
\end{table}

\subsection{Result}
Multiple metrics are employed to measure the speech enhancement performance, including perceptual evaluation speech quality (PESQ)\cite{rix2001perceptual}, short-time objective intelligibility (STOI)\cite{taal2011algorithm} and word error rate (WER). A pretrained model based on the conformer architectures\cite{wu2021u2++} are used to compute the WER.

We compare MNTFA with DCCRN and DPCRN on the test set. From the comparison results shown in Table \ref{tbl:table1}, our approach produces a superior performance to DCCRN and DPCRN in terms of PESQ and STOI. In table \ref{tbl:table2}, "Para.", "MACs", and "WER" respectively represent the parameter amount of the model, computation complexity, and  word error rate for ASR performance. With 0.23 M parameters, 1.89 Giga floating-point operations per second (GFLOPs) and without any future information, our model outperform DPCRN with 0.8M parameters.

\begin{table}[htbp]
\caption{Ablation study on test set.}
\centering
\begin{tabular}{ccccc}
\hline
System & Loss                             & PESQ & STOI  & WER   \\ \hline
Noisy  &      -                            & 1.85 & 72.76 & 13.13 \\
MNTFA  & $L_{MSE}$                       & 2.67 & 81.15 & 13.79 \\
MNTFA  & $L_{MSE}+L_{ASR}$            & 2.66 & 81.27 & 12.74 \\
MNTFA  & $L_{MSE}+L_{aux}$            &\textbf{2.77} & \textbf{81.64} & 14.35 \\
MNTFA  & $L_{MSE}+L_{ASR}+L_{aux}$ & 2.72 & 81.46 & \textbf{12.55} \\ \hline
\end{tabular}
\label{tbl:table3}
\end{table}

At the same time, we conducted ablation experiments on the proposed model to show the importance of the different losses. Table \ref{tbl:table3} shows the ablation results. After removing ASR loss, PESQ, STOI, WER increased by 0.05, 0.18, and 1.8, respectively. After removing multi-resolution STFT auxiliary loss, PESQ and STOI decreased by 0.06 and 0.37, respectively, but WER increased 0.19.  The results show that MNTFA with MSE loss and multi-resolution STFT auxiliary loss improved the performance in PESQ and STOI, but not the WER. We found that there is no absolute direct correlation between PESQ and WER, as a higher PESQ score does not necessarily guarantee a lower WER, which means the higher PESQ should not always be the final goal.

\section*{CONCLUSION}
In this paper, we propose a multi-loss network with Time-Frequency attention for speech enhancement.
In the proposed method, the self-attention blocks are utilized to model the long-time relations, and we exploit the T-attention and F-attention to model the temporal and frequency decencies. With 0.23 M parameters, our model achieves competitive results and we also found that there is no absolute direct correlation between PESQ and WER, as a higher PESQ score does not necessarily guarantee a lower WER. In the future, we will investigate a better way to ensure that speech recognition accuracy while also preserving a good noise reduction performance.

\bibliographystyle{IEEEtran}
\bibliography{IEEEexample}

\begin{thebibliography}{10}
\providecommand{\url}[1]{#1}
\csname url@samestyle\endcsname
\providecommand{\newblock}{\relax}
\providecommand{\bibinfo}[2]{#2}
\providecommand{\BIBentrySTDinterwordspacing}{\spaceskip=0pt\relax}
\providecommand{\BIBentryALTinterwordstretchfactor}{4}
\providecommand{\BIBentryALTinterwordspacing}{\spaceskip=\fontdimen2\font plus
\BIBentryALTinterwordstretchfactor\fontdimen3\font minus
  \fontdimen4\font\relax}
\providecommand{\BIBforeignlanguage}[2]{{%
\expandafter\ifx\csname l@#1\endcsname\relax
\typeout{** WARNING: IEEEtran.bst: No hyphenation pattern has been}%
\typeout{** loaded for the language `#1'. Using the pattern for}%
\typeout{** the default language instead.}%
\else
\language=\csname l@#1\endcsname
\fi
#2}}
\providecommand{\BIBdecl}{\relax}
\BIBdecl

\bibitem{20061486}
\BIBentryALTinterwordspacing
S.~Srinivasan, N.~Roman, and D.~Wang, ``Binary and ratio time-frequency masks
  for robust speech recognition,'' \emph{Speech Communication}, vol.~48,
  no.~11, pp. 1486--1501, 2006, robustness Issues for Conversational
  Interaction. [Online]. Available:
  \url{https://www.sciencedirect.com/science/article/pii/S0167639306001129}
\BIBentrySTDinterwordspacing

\bibitem{6639038}
A.~Narayanan and D.~Wang, ``Ideal ratio mask estimation using deep neural
  networks for robust speech recognition,'' in \emph{2013 IEEE International
  Conference on Acoustics, Speech and Signal Processing}, 2013, pp. 7092--7096.

\bibitem{7472934}
Y.~Zhao, D.~Wang, I.~Merks, and T.~Zhang, ``Dnn-based enhancement of noisy and
  reverberant speech,'' in \emph{2016 IEEE International Conference on
  Acoustics, Speech and Signal Processing (ICASSP)}, 2016, pp. 6525--6529.

\bibitem{6665000}
Y.~Xu, J.~Du, L.-R. Dai, and C.-H. Lee, ``An experimental study on speech
  enhancement based on deep neural networks,'' \emph{IEEE Signal Processing
  Letters}, vol.~21, no.~1, pp. 65--68, 2014.

\bibitem{Yin_Luo_Xiong_Zeng_2020}
\BIBentryALTinterwordspacing
D.~Yin, C.~Luo, Z.~Xiong, and W.~Zeng, ``Phasen: A phase-and-harmonics-aware
  speech enhancement network,'' \emph{Proceedings of the AAAI Conference on
  Artificial Intelligence}, vol.~34, no.~05, pp. 9458--9465, Apr. 2020.
  [Online]. Available:
  \url{https://ojs.aaai.org/index.php/AAAI/article/view/6489}
\BIBentrySTDinterwordspacing

\bibitem{fu2018end}
S.-W. Fu, T.-W. Wang, Y.~Tsao, X.~Lu, and H.~Kawai, ``End-to-end waveform
  utterance enhancement for direct evaluation metrics optimization by fully
  convolutional neural networks,'' \emph{IEEE/ACM Transactions on Audio,
  Speech, and Language Processing}, vol.~26, no.~9, pp. 1570--1584, 2018.

\bibitem{Wave-U-Net}
S.~Daniel, E.~Sebastian, and D.~Simon, ``Wave-u-net: A multi-scale neural
  network for end-to-end audio source separation,'' \emph{arXiv preprint
  arXiv:1806.03185}, 2018.

\bibitem{zhang2020furcanext}
L.~Zhang, Z.~Shi, J.~Han, A.~Shi, and D.~Ma, ``Furcanext: End-to-end monaural
  speech separation with dynamic gated dilated temporal convolutional
  networks,'' in \emph{MultiMedia Modeling: 26th International Conference, MMM
  2020, Daejeon, South Korea, January 5--8, 2020, Proceedings, Part I
  26}.\hskip 1em plus 0.5em minus 0.4em\relax Springer, 2020, pp. 653--665.

\bibitem{luo2019conv-tasnet}
Y.~Luo and N.~Mesgarani, ``Conv-tasnet: Surpassing ideal time-frequency
  magnitude masking for speech separation,'' \emph{IEEE Transactions on Audio,
  Speech, and Language Processing}, pp. 1--1, 2019.

\bibitem{wang2005ideal}
D.~Wang, ``On ideal binary mask as the computational goal of auditory scene
  analysis,'' \emph{Speech separation by humans and machines}, pp. 181--197,
  2005.

\bibitem{wang2014training}
Y.~Wang, A.~Narayanan, and D.~Wang, ``On training targets for supervised speech
  separation,'' \emph{IEEE/ACM transactions on audio, speech, and language
  processing}, vol.~22, no.~12, pp. 1849--1858, 2014.

\bibitem{8682834}
K.~Tan and D.~Wang, ``Complex spectral mapping with a convolutional recurrent
  network for monaural speech enhancement,'' in \emph{ICASSP 2019 - 2019 IEEE
  International Conference on Acoustics, Speech and Signal Processing
  (ICASSP)}, 2019, pp. 6865--6869.

\bibitem{le2021dpcrn}
X.~Le, H.~Chen, K.~Chen, and J.~Lu, ``Dpcrn: Dual-path convolution recurrent
  network for single channel speech enhancement,'' \emph{arXiv preprint
  arXiv:2107.05429}, 2021.

\bibitem{luo2020dual}
Y.~Luo, Z.~Chen, and T.~Yoshioka, ``Dual-path rnn: efficient long sequence
  modeling for time-domain single-channel speech separation,'' in \emph{ICASSP
  2020-2020 IEEE International Conference on Acoustics, Speech and Signal
  Processing (ICASSP)}.\hskip 1em plus 0.5em minus 0.4em\relax IEEE, 2020, pp.
  46--50.

\bibitem{crn2018}
T.~Ke and W.~DeLiang, ``A convolutional recurrent neural network for real-time
  speech enhancement,'' \emph{Interspeech}, pp. 3229--3233, 2018.

\bibitem{zhang2022multi}
G.~Zhang, C.~Wang, L.~Yu, and J.~Wei, ``Multi-scale temporal frequency
  convolutional network with axial attention for multi-channel speech
  enhancement,'' in \emph{ICASSP 2022-2022 IEEE International Conference on
  Acoustics, Speech and Signal Processing (ICASSP)}.\hskip 1em plus 0.5em minus
  0.4em\relax IEEE, 2022, pp. 9206--9210.

\bibitem{yamamoto2020parallel}
R.~Yamamoto, E.~Song, and J.-M. Kim, ``Parallel wavegan: A fast waveform
  generation model based on generative adversarial networks with
  multi-resolution spectrogram,'' in \emph{ICASSP 2020-2020 IEEE International
  Conference on Acoustics, Speech and Signal Processing (ICASSP)}.\hskip 1em
  plus 0.5em minus 0.4em\relax IEEE, 2020, pp. 6199--6203.

\bibitem{hsieh2020improving}
T.-A. Hsieh, C.~Yu, S.-W. Fu, X.~Lu, and Y.~Tsao, ``Improving perceptual
  quality by phone-fortified perceptual loss using wasserstein distance for
  speech enhancement,'' \emph{arXiv preprint arXiv:2010.15174}, 2020.

\bibitem{chen2022wavlm}
S.~Chen, C.~Wang, Z.~Chen, Y.~Wu, S.~Liu, Z.~Chen, J.~Li, N.~Kanda,
  T.~Yoshioka, X.~Xiao \emph{et~al.}, ``Wavlm: Large-scale self-supervised
  pre-training for full stack speech processing,'' \emph{IEEE Journal of
  Selected Topics in Signal Processing}, vol.~16, no.~6, pp. 1505--1518, 2022.

\bibitem{wang2018supervised}
D.~Wang and J.~Chen, ``Supervised speech separation based on deep learning: An
  overview,'' \emph{IEEE/ACM Transactions on Audio, Speech, and Language
  Processing}, vol.~26, no.~10, pp. 1702--1726, 2018.

\bibitem{hu2020dccrn}
Y.~Hu, Y.~Liu, S.~Lv, M.~Xing, S.~Zhang, Y.~Fu, J.~Wu, B.~Zhang, and L.~Xie,
  ``Dccrn: Deep complex convolution recurrent network for phase-aware speech
  enhancement,'' \emph{arXiv preprint arXiv:2008.00264}, 2020.

\bibitem{he2016deep}
K.~He, X.~Zhang, S.~Ren, and J.~Sun, ``Deep residual learning for image
  recognition,'' in \emph{Proceedings of the IEEE conference on computer vision
  and pattern recognition}, 2016, pp. 770--778.

\bibitem{he2015delving}
------, \emph{Delving deep into rectifiers: Surpassing human-level performance
  on imagenet classification}.\hskip 1em plus 0.5em minus 0.4em\relax
  Proceedings of the IEEE international conference on computer vision
  1026--1034, 2015.

\bibitem{yamamoto2019probability}
R.~Yamamoto, E.~Song, and J.-M. Kim, ``Probability density distillation with
  generative adversarial networks for high-quality parallel waveform
  generation,'' \emph{arXiv preprint arXiv:1904.04472}, 2019.

\bibitem{arik2018fast}
S.~{\"O}. Ar{\i}k, H.~Jun, and G.~Diamos, ``Fast spectrogram inversion using
  multi-head convolutional neural networks,'' \emph{IEEE Signal Processing
  Letters}, vol.~26, no.~1, pp. 94--98, 2018.

\bibitem{reddy2020interspeech}
C.~K. Reddy, V.~Gopal, R.~Cutler, E.~Beyrami, R.~Cheng, H.~Dubey,
  S.~Matusevych, R.~Aichner, A.~Aazami, S.~Braun \emph{et~al.}, ``The
  interspeech 2020 deep noise suppression challenge: Datasets, subjective
  testing framework, and challenge results,'' \emph{arXiv preprint
  arXiv:2005.13981}, 2020.

\bibitem{rix2001perceptual}
A.~W. Rix, J.~G. Beerends, M.~P. Hollier, and A.~P. Hekstra, ``Perceptual
  evaluation of speech quality (pesq)-a new method for speech quality
  assessment of telephone networks and codecs,'' in \emph{2001 IEEE
  international conference on acoustics, speech, and signal processing.
  Proceedings (Cat. No. 01CH37221)}, vol.~2.\hskip 1em plus 0.5em minus
  0.4em\relax IEEE, 2001, pp. 749--752.

\bibitem{taal2011algorithm}
C.~H. Taal, R.~C. Hendriks, R.~Heusdens, and J.~Jensen, ``An algorithm for
  intelligibility prediction of time--frequency weighted noisy speech,''
  \emph{IEEE Transactions on Audio, Speech, and Language Processing}, vol.~19,
  no.~7, pp. 2125--2136, 2011.

\bibitem{wu2021u2++}
D.~Wu, B.~Zhang, C.~Yang, Z.~Peng, W.~Xia, X.~Chen, and X.~Lei, ``U2++: Unified
  two-pass bidirectional end-to-end model for speech recognition,'' \emph{arXiv
  preprint arXiv:2106.05642}, 2021.

\end{thebibliography}

\end{document}